\documentclass[twocolumn,superscriptaddress,showpacs,preprintnumbers,prl]{revtex4} 

\usepackage{graphicx}
\usepackage{amsmath} 
\usepackage{times}
\usepackage{epsfig}
\usepackage{color}
\usepackage{ulem}   
\begin{document}

\def\salto{\vskip 1cm} \def\lag{\langle} \def\rag{\rangle}

\newcommand{\redit}[1]{\textcolor{red}{#1}}
\newcommand{\blueit}[1]{\textcolor{blue}{#1}}
\newcommand{\magit}[1]{\textcolor{magenta}{#1}}

\newcommand{\MSTD} {Materials Science and Technology Division, Oak Ridge
National Laboratory, Oak Ridge, TN 37831, USA}
\newcommand{\CNMS} {Center for Nanophase Materials Sciences, Oak Ridge National
 Laboratory, Oak Ridge, TN 37831, USA}
\newcommand{\LLNL} {Lawrence Livermore National Laboratory, Livermore, CA
94550, USA}

\title{Quantum Monte Carlo Calculations of Dihydrogen Binding Energetics on Ca Cations: an Assessment of Errors in Density Functionals for Weakly Bonded Systems}

\author{Michal Bajdich}           \affiliation {\MSTD}
\author{Fernando A. Reboredo}       \affiliation {\MSTD}
\author{P. R. C. Kent}        \affiliation {\CNMS}
  
\begin{abstract}
  We investigate the binding of single and quadruple hydrogen
  molecules on a positively charged Ca ion. By comparing with
  benchmark quantum Monte Carlo (QMC) calculations we demonstrate wide
  variability in other more approximate electronic structure methods
  including common density functionals. Single determinant QMC calculations find no
  binding at short range by approximately 0.1 eV for the quadruple
  hydrogen molecule case, for a fixed hydrogen bond length of 0.77
  Angstrom. Density functional calculations using common functionals
  such a LDA and B3LYP  differ substantially from the QMC binding
  curve.  We show that use of full Hartree--Fock exchange and PBE correlation
  (HFX+PBEC) obtains close agreement with the QMC results, both
  qualitatively and quantitatively.  These results both motivate the
  use and development of improved functionals and indicate that
  caution is required applying electronic structure methods to weakly
  bound systems such as hydrogen storage materials based on metal ion
  decorated nanostructures.
\end{abstract}

\date{\today}

\maketitle

Our ability to accurately predict molecular adsorption energies is of
widespread importance in the physical, chemical, and materials
sciences. Technologically, the adsorption of small molecules on
semiconductors or metals is an essential step in many catalytic or
energy storage related areas. In the case of hydrogen storage, the
strength of the adsorption can determine the suitability a material
for practical application: if the binding is too high, release of the
hydrogen will be difficult at moderate operating temperatures, while
if the binding is too weak, storage of the hydrogen
will be ineffective.

Calculating the energetics of hydrogen adsorption is a difficult task
and requires highly accurate quantum mechanics based calculations. If the
structure and eventually the dynamics of the adsorption process are to
be accurately modeled, the potential energy surface of the adsorbant
and adsorbate must be accurately simulated over a length scale of at
least 5 Angstrom. Density functional theory (DFT) based methods are
the most widely applied electronic structure methods for studies
of hydrogen storage materials. However, in practice the DFTs
are not only approximate, but are also rarely benchmarked in the
non-bonding and weakly bonding configurations vital for hydrogen storage.

Motivated in part by recent discussions and discrepancies for DFT
predictions of hydrogen adsorption on alkaline earth
metals\cite{cha_inaccuracy_2009,ohk_commentinaccuracy_2010,cha_cha_2010},
we have performed extensive Quantum Monte Carlo
(QMC)\cite{foulkes_quantum_2001} calculations for H$_2$ (dihydrogen)
adsorption on the Ca$^{1+}$ system. The ion's charge models the
scenario where the ion is absorbed on
graphene\cite{ataca_hydrogen_2009,kim_crossover_2009}.  QMC provides
an accurate and unbiased reference to compare against approximate but
more computationally affordable approaches. We concentrate on the
interaction of hydrogens with a single ion, as opposed to a system
with a substrate, since the geometries are easily and unambiguously
specified and the systems are already sufficient to demonstrate
substantial differences in predicted binding energies and overall
shape of the binding curves.

The interaction of one or many H$_2$ molecules with Ca$^{1+}$ will
undoubtedly involve several effects: charge transfer, polarization,
and potential long range dispersion (or van der Waals)
interactions. To accurately model these systems, first principles
calculations should be able to accurately account for all these
effects with little reliance on, e.g. error cancelation.  For example,
van der Waals interactions interactions are naturally and accurately
included within QMC
approaches\cite{sorella_weaK_2007,beaudet_molecular_2008,santra_accuracy_2008,ma_water-benzene_2009},
but are absent from common DFTs.

In the following we (i) describe our QMC 
methodology, (ii) present benchmark results for the cases of 
single H$_2$ adsorption on Ca$^{1+}$, and (iii) since in actual use scenarios
additional hydrogen molecules will be present we also
results for quadruple H$_2$ adsorption. These systems are constructed
identically to those of Ref. \cite{cha_inaccuracy_2009}. Finally, (iv) we summarize our findings.

\textit{Quantum Monte Carlo}--- The QMC method allows for a
very efficient and accurate solution of the Schr\"odinger equation. In
contrast with many electronic structure methods, QMC methods involve
only well controlled approximations. Although their computational
prefactor is often large, for small and medium sized molecular systems
energetics close to chemical accuracy can be obtained,
e.g. Refs. \cite{umrigar_alleviation_2007,bajdich_systematic_2010}.
These properties make QMC methods ideal for benchmark studies and for the
assessment of computationally cheaper but more approximate methods.

QMC methods are wavefunction based and the most important input is the
trial many body wavefunction. In Variational Monte Carlo (VMC) a
direct variational evaluation of the energy of a trial wavefunction is
performed using importance sampled Monte Carlo integration. VMC
calculations therefore suffer from a potentially very strong
dependence on the input wave function and any prior assumptions about
the electronic structure, but have the advantage that the actual
many-body wavefunction is obtained and can be analyzed.  In fixed-node
diffusion quantum Monte Carlo (DMC), the lowest energy state
consistent with the zeros (nodes) of the trial wavefunction is
projected. This projection greatly reduces the dependence of the final
energy on the input trial wavefunction compared to VMC. In practice,
very accurate results are obtained by DMC for a wide variety of
molecular and solid state
systems\cite{foulkes_quantum_2001,grossman_benchmark_2002,sorella_weaK_2007,beaudet_molecular_2008,santra_accuracy_2008,ma_water-benzene_2009,nemec_benchmark_2010}. Due
to the increased robustness we concentrate on DMC results in this
study.

For our purposes, the only significant approximations in DMC
calculations are (i) the use of pseudopotentials and (ii) the
fixed-node approximation and consequent dependence on the nodal
surface of the input trial wavefunction. The first approximation
introduces systematic errors via the approximate treatment of
core-valence interactions and via the locality
approximation\cite{mitas_nonlocal_1991} necessary to evaluate the non-local
pseudopotentials in DMC. We minimize these errors by using a small Ne
core for the Ca pseudopotential
\cite{burkatzki_energy-consistent_2007} and very high quality trial
wavefunctions.  We use the same
pseudopotentials in all our calculations to ensure a fair comparison
between all methods: the same Hamiltonian is solved in our QMC, DFT, and
quantum chemical calculations.

To minimize the nodal errors in our DMC calculations we also use multideterminant trial
wavefunctions obtained from configuration interaction calculations
that are subsequently reoptimized via the energy minimization
method\cite{umrigar_alleviation_2007}. This approach is a significant
advance over conventional applications of DMC where the nodal surface
of the trial wavefunction consists of only a single Slater determinant
determined by a less accurate theory such as DFT: the nodal errors are
systematically reduced to near chemical
accuracy\cite{umrigar_alleviation_2007,bajdich_systematic_2010} when
sufficient statistics can be obtained and the multideterminant
expansion is large enough. Previous studies have shown that (i) for
light molecules single determinant DMC yields results similar in
accuracy CCSD(T) with the aug-cc-pVQZ basis
set\cite{grossman_benchmark_2002}, (ii) these errors are further
reduced with multideterminant methods,
e.g. \cite{grossman_benchmark_2002,umrigar_alleviation_2007,bajdich_systematic_2010},
and (iii) pseudopotential errors are small and less significant than
the nodal error in these
systems\cite{grossman_benchmark_2002,nemec_benchmark_2010}.

In principle, modern trial wavefunction optimization
methods\cite{umrigar_alleviation_2007,bajdich_systematic_2010} can
produce DMC results nearly independent of the input provided
sufficiently flexible trial wavefunction forms are adopted. Here we
validate our single determinant nodal surface results using large
configuration interaction expansions of many determinants.

In the following calculations we use trial wavefunctions consisting of
a weighted sum of Slater determinants multiplied by a two-body Jastrow
factor. The Slater determinants consist of orbitals determined by
GAMESS\cite{gamess09} DFT or complete active space self consistent
field (CASSCF) calculations expanded in the large ANO-VTZ gaussian basis set\cite{burkatzki_energy-consistent_2007}. The
two-body Jastrow factor does not change the nodal surface but acts to
enforce the electron-electron cusp condition, greatly improving the
overall quality of the trial wavefunctions. For the QMC calculations
we used the QWALK code\cite{wagner_qwalk_2009}. Multideterminant QMC calculations
used up to 370 determinants, where we took all CASSCF determinant of
squared magnitude greater than 0.01.  Energy
minimization was performed starting from the truncated CASSCF results.
We note that the DMC energies always lie substantially below the pure
quantum chemistry results.  Pseudopotentials were derived in the soft Hartree-Fock
formalism\cite{burkatzki_energy-consistent_2007}. An average DMC
population of $\sim 30000$ walkers and a small time step of 0.005 a.u. was
used. The largest DMC calculations used O(1000) processor hours per
energy point.

\begin{figure}
\includegraphics[width=0.95\linewidth,clip=true]{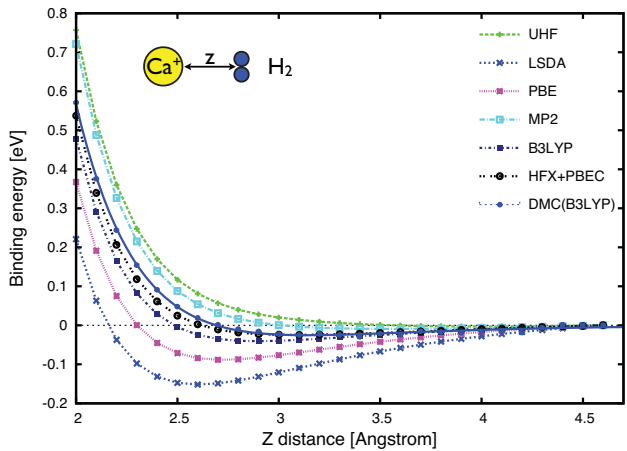}
\caption{Calculated binding energy for a single hydrogen 
  molecule (H$_2$) approaching Ca$^{1+}$. The molecule to oriented with the
  bond perpendicular to the line of approach (see inset). Results are shown for
  the unrestricted Hartree--Fock (UHF), the local spin density
  approximation (LSDA), PBE and B3LYP density functionals,
  second order M{\o}ller--Plesset perturbation theory (MP2). We also show Diffusion
  Quantum Monte Carlo (DMC) and density functional results calculated
  using exact exchange combined with PBE correlation (HFX+PBEC). The DMC calculations use a single
  determinant of  B3LYP orbitals and a Jastrow factor for the trial
  wavefunction and nodal surface. The hydrogen
  molecule bond length is held fixed at 0.77 Angstrom. The lines
are a guide to the eye. Error bars are smaller than the
size of the DMC symbols.
\label{fig:1h2}}
\end{figure}

\textit{Results for single H$_2$ absorption}--- Figure \ref{fig:1h2}
shows our calculated binding energy curve for the hydrogen dimer on
Ca$^{1+}$, with the molecular bond oriented perpendicularly to the
line of approach. Since the calculation of forces is not well
developed in QMC, for each distance from the Ca$^{1+}$ ion we computed
energies for all methods with a fixed bond length of 0.77 Angstrom,
corresponding to the value found near binding in quantum chemical
calculations \cite{cha_inaccuracy_2009}. Only small changes in the
fully relaxed value are seen over all distances, indicating that the
trends in the binding at  fixed bond length are representative of the 
relaxed case. For computational simplicity, we treat the energy of the
system at $z=4.6$ Angstrom as representing fully separated unbound system.

Our results show that while the potential energy surface varies
quantitatively between the methods, for a single dimer the general
trends given are qualitatively similar for many of the methods, with a
single minimum. However, unrestricted Hartree--Fock (UHF) and
second order {M}{\o}ller--Plesset perturbation theory (MP2) calculations
show negligible binding. The DMC data shows a minimum
around $z=3.1$ Angstrom, and binding of $\sim 0.025$ eV.

Comparing the density functional results against the DMC energy curve
we find that B3LYP\cite{becke_density-functional_1988} functional
gives a relatively good agreement, with minimum at $z=<2.9$ Angstrom.
However, the LSDA functional\cite{perdew_self-interaction_1981}
significantly overbinds by at least 0.1 eV, while the PBE
functional\cite{perdew_generalized_1996} lies midway between the B3LYP
and LSDA values. None of the functionals results in false energetic
minima in the binding energy curve, however the distance of the
minimum energy varies by 0.5 Angstrom over these
functionals. Calculations using Hartree--Fock exchange combined with PBE
correlation
(HFX+PBEC) (similar to
Refs. \onlinecite{ernzerhof_assessment_1999,adamo_toward_1999} except
with 100\% exchange)  very
closely resemble the DMC results; analysis and possible reasons
for the apparent accuracy are discussed after the four dihydrogen results.

\begin{figure*}
\includegraphics[width=0.60\linewidth,clip=true]{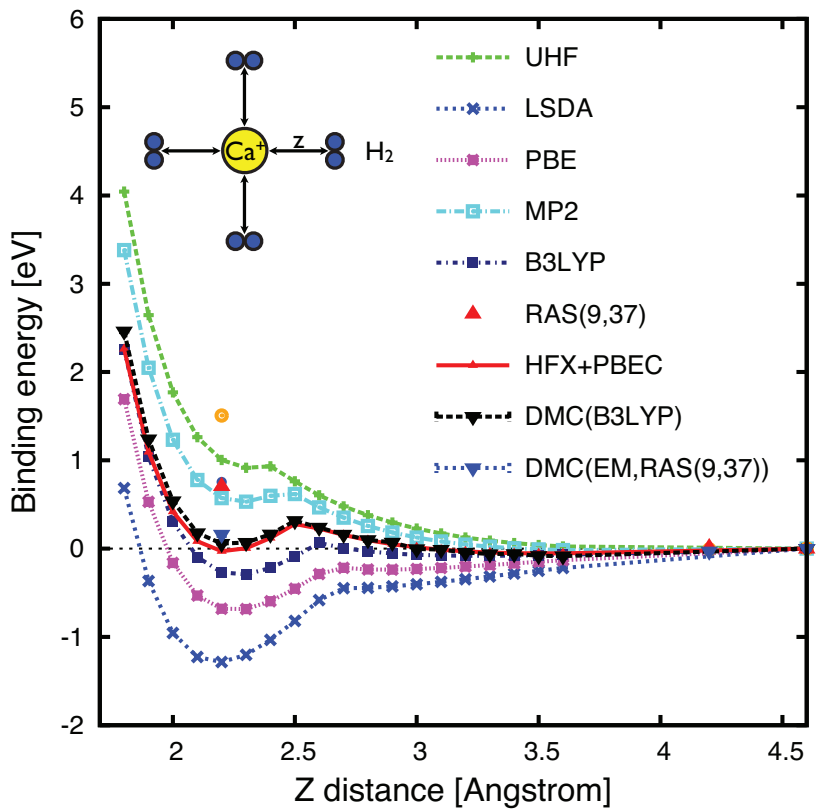}
\caption{Calculated binding energy for a four dihydrogen (H$_2$)
  molecules approaching Ca$^{1+}$. Each molecule is oriented with the
  bond perpendicular to the line of approach in an overall planar
  geometry with D4 symmetry (see inset). Results are shown for
  unrestricted Hartree-Fock (HF), density functional theory using
  several approximate functionals (LDA, PBE, B3LYP), and for two sets
  of Diffusion Quantum Monte Carlo (DMC) calculations. For the range
  of distances studied, DMC data is given for a single determinant of
  B3LYP orbitals and a Jastrow factor for the trial wavefunction and
  nodal surface (black triangles). At 2.2 Angstrom separation we also
  compute the binding energy in DMC using Energy Minimized Restricted
  Active Space (RAS) multideterminant wavefunctions. At this distance
  we also show quantum chemical  results for RAS(9,37) (red triangle), RAS (9,31) (blue circle), and
  complete active space CAS(9,18) (orange circle).  HFX+PBEC
  indicates density functional results calculated using exact exchange
  combined with PBE correlation. The hydrogen
  molecule bond lengths are held fixed at 0.77 Angstrom. See text for
  more information. The lines drawn are a guide to the eye. DMC error bars
  are smaller than the symbols. 
  \label{fig:4h2}}
\end{figure*}

\textit{Results for four H$_2$ absorption}--- Figure \ref{fig:4h2}
shows our calculated binding energy curve for four hydrogen dimers on
Ca$^{1+}$. In this system the hydrogens are pinned in a planar
geometry, ninety degrees apart in $D_{4}$ symmetry, with molecular bonds
oriented perpendicularly to the line of approach (inset in Figure
\ref{fig:4h2}).

The binding energies obtained with single determinant DMC display a
minimum around 2.2 Angstrom. However, comparing these energies with
those over 3 Angstrom clearly shows the minimum to be a local
metastable minimum: there is no binding of four H$_2$ molecules in
this planar geometry at short range at the single determinant DMC
level. To test the accuracy of these calculations we also used
multideterminant wavefunctions determinants, initially obtained by
restricted active space RAS(9,37) calculations.  The binding is
shifted to higher energies by $\sim 0.1$ eV indicating that the single
determinant results and nodal surface are robust.

DFT based calculations show clear energy minima around the 2.2
Angstrom distance indicating significant binding of the H$_2$
molecules. Although the depth of binding varies, similar behavior is
obtained for LSDA, PBE and B3LYP. The same energetic ordering is
observed as for the single hydrogen case, with LSDA displaying
greatest binding. By contrast, UHF calculations display only a slight
minimum around 2.3 Angstrom.  We also include quantum chemical results
from RAS and complete active space (CAS) calculations in
Fig.\ref{fig:4h2}. The larger active space calculations reduce the
calculated binding energy, moving the quantum chemical results towards
the DMC results. Perturbative theory results (MP2) also show no overall
binding. A clear transition between states of $A_{1g}$ and $B_{2g}$
symmetry is observed\cite{cha_inaccuracy_2009} between 2.5 and 3.0
Angstrom depending on the underlying theory.

Our qualitative conclusion of no binding for the four hydrogen
molecule case with fixed 0.77 Angstrom bond length is in qualitative
agreement with previous quantum chemical
results\cite{cha_inaccuracy_2009}. However our more extensive basis
sets and more rigorous DMC calculations reveal that that the energy
scale for any potential binding is very small, only a few tenths of an
eV. Strikingly, even allowing a generous estimate of 0.2eV residual
systematic errors in our DMC calculations, any eventual binding will
remain small (order 0.1eV) whereas the LSDA and PBE functionals
predict binding energies one order of magnitude larger.

We also tested energies obtained from density functional theory using
using Hartree--Fock exchange combined with PBE correlation, gradually
increasing the fraction of exchange.  As for the single hydrogen
molecule case, the calculated binding curve accurately follows the DMC
data over all distances.  However, only a 100\% contribution fully
reproduced the DMC data; lesser contributions smoothly interpolating
between the DMC and PBE results.  Given the wide variation seen for
other functionals, this is noteworthy, and also indicates the primary
source of error in the other density functional predictions: We argue,
in accordance with Ref. \cite{kummel_orbital-dependent_2008}, that the
semi-local functionals provide good description of static correlation
and exchange in this system. On the other hand, dynamic or
long-range exchange is mostly absent in these functionals. Since our
system seems not to have a strong multi-reference character (based on
our RAS and CAS calculations) the dynamical part of exchange must play
a dominant role.  This last part is well described only in HF or in
exact-exchange functionals such as optimized effective potential (OEP)
method. Therefore including full dynamical part of a exchange together
with the correlation from LSDA, PBE, or B3LYP gives a good
description. This observation is mostly independent from the type of
the correlation used. 

\textit{Summary}--- We have performed benchmark quantum Monte Carlo
calculations of hydrogen molecule binding on the Ca cation. Density
functional calculations vary widely in the predicted binding
energy. Density functional calculations using Hartree--Fock exchange well reproduce the Monte Carlo
results, suggesting a route -- if not universal -- to predictive and accurate calculations in
this and related systems. We hope that our results will help further motivate the
development of improved functionals. As metal ions have been proposed
as hydrogen binding centers in new hydrogen storage materials, we
strongly recommend caution in applying density functional methods to
these systems. Appropriate benchmarking using quantum chemical or
quantum Monte Carlo techniques is required.

We thank V. R. Cooper for helpful conversations.  This research used
computer resources supported by the U.S. DOE Office of Science under
contract DE-AC02-05CH11231 (NERSC) and DE-AC05-00OR22725
(NCCS). Research sponsored by U.S. DOE BES Division of Materials
Sciences \& Engineering (FAR) and ORNL LDRD program (MB). The Center
for Nanophase Materials Sciences research was sponsored by the
U. S. DOE Division of Scientific User Facilities (PRCK).


\begin{thebibliography}{24}
\expandafter\ifx\csname natexlab\endcsname\relax\def\natexlab#1{#1}\fi
\expandafter\ifx\csname bibnamefont\endcsname\relax
  \def\bibnamefont#1{#1}\fi
\expandafter\ifx\csname bibfnamefont\endcsname\relax
  \def\bibfnamefont#1{#1}\fi
\expandafter\ifx\csname citenamefont\endcsname\relax
  \def\citenamefont#1{#1}\fi
\expandafter\ifx\csname url\endcsname\relax
  \def\url#1{\texttt{#1}}\fi
\expandafter\ifx\csname urlprefix\endcsname\relax\def\urlprefix{URL }\fi
\providecommand{\bibinfo}[2]{#2}
\providecommand{\eprint}[2][]{\url{#2}}

\bibitem[{\citenamefont{Cha et~al.}(2009)\citenamefont{Cha, Lim, Choi, Cha, and
  Park}}]{cha_inaccuracy_2009}
\bibinfo{author}{\bibfnamefont{J.}~\bibnamefont{Cha}},
  \bibinfo{author}{\bibfnamefont{S.}~\bibnamefont{Lim}},
  \bibinfo{author}{\bibfnamefont{C.~H.} \bibnamefont{Choi}},
  \bibinfo{author}{\bibfnamefont{M.}~\bibnamefont{Cha}}, \bibnamefont{and}
  \bibinfo{author}{\bibfnamefont{N.}~\bibnamefont{Park}},
  \bibinfo{journal}{Physical Review Letters} \textbf{\bibinfo{volume}{103}},
  \bibinfo{pages}{216102} (\bibinfo{year}{2009}).

\bibitem[{\citenamefont{Ohk et~al.}(2010)\citenamefont{Ohk, Kim, and
  Jung}}]{ohk_commentinaccuracy_2010}
\bibinfo{author}{\bibfnamefont{Y.}~\bibnamefont{Ohk}},
  \bibinfo{author}{\bibfnamefont{Y.}~\bibnamefont{Kim}}, \bibnamefont{and}
  \bibinfo{author}{\bibfnamefont{Y.}~\bibnamefont{Jung}},
  \bibinfo{journal}{Physical Review Letters} \textbf{\bibinfo{volume}{104}},
  \bibinfo{pages}{179601} (\bibinfo{year}{2010}).

\bibitem[{\citenamefont{Cha et~al.}(2010)\citenamefont{Cha, Choi, and
  Park}}]{cha_cha_2010}
\bibinfo{author}{\bibfnamefont{J.}~\bibnamefont{Cha}},
  \bibinfo{author}{\bibfnamefont{C.~H.} \bibnamefont{Choi}}, \bibnamefont{and}
  \bibinfo{author}{\bibfnamefont{N.}~\bibnamefont{Park}},
  \bibinfo{journal}{Physical Review Letters} \textbf{\bibinfo{volume}{104}},
  \bibinfo{pages}{179602} (\bibinfo{year}{2010}).

\bibitem[{\citenamefont{Foulkes et~al.}(2001)\citenamefont{Foulkes, Mitas,
  Needs, and Rajagopal}}]{foulkes_quantum_2001}
\bibinfo{author}{\bibfnamefont{W.~M.~C.} \bibnamefont{Foulkes}},
  \bibinfo{author}{\bibfnamefont{L.}~\bibnamefont{Mitas}},
  \bibinfo{author}{\bibfnamefont{R.~J.} \bibnamefont{Needs}}, \bibnamefont{and}
  \bibinfo{author}{\bibfnamefont{G.}~\bibnamefont{Rajagopal}},
  \bibinfo{journal}{Reviews of Modern Physics} \textbf{\bibinfo{volume}{73}},
  \bibinfo{pages}{33} (\bibinfo{year}{2001}).

\bibitem[{\citenamefont{Ataca et~al.}(2009)\citenamefont{Ataca, Akt�rk, and
  Ciraci}}]{ataca_hydrogen_2009}
\bibinfo{author}{\bibfnamefont{C.}~\bibnamefont{Ataca}},
  \bibinfo{author}{\bibfnamefont{E.}~\bibnamefont{Akt�rk}}, \bibnamefont{and}
  \bibinfo{author}{\bibfnamefont{S.}~\bibnamefont{Ciraci}},
  \bibinfo{journal}{Physical Review B} \textbf{\bibinfo{volume}{79}},
  \bibinfo{pages}{041406} (\bibinfo{year}{2009}).

\bibitem[{\citenamefont{Kim et~al.}(2009)\citenamefont{Kim, Jhi, Lim, and
  Park}}]{kim_crossover_2009}
\bibinfo{author}{\bibfnamefont{G.}~\bibnamefont{Kim}},
  \bibinfo{author}{\bibfnamefont{S.}~\bibnamefont{Jhi}},
  \bibinfo{author}{\bibfnamefont{S.}~\bibnamefont{Lim}}, \bibnamefont{and}
  \bibinfo{author}{\bibfnamefont{N.}~\bibnamefont{Park}},
  \bibinfo{journal}{Physical Review B} \textbf{\bibinfo{volume}{79}},
  \bibinfo{pages}{155437} (\bibinfo{year}{2009}).

\bibitem[{\citenamefont{Sorella et~al.}(2007)\citenamefont{Sorella, Casula, and
  Rocca}}]{sorella_weaK_2007}
\bibinfo{author}{\bibfnamefont{S.}~\bibnamefont{Sorella}},
  \bibinfo{author}{\bibfnamefont{M.}~\bibnamefont{Casula}}, \bibnamefont{and}
  \bibinfo{author}{\bibfnamefont{D.}~\bibnamefont{Rocca}},
  \bibinfo{journal}{Journal of Chemical Physics}
  \textbf{\bibinfo{volume}{127}}, \bibinfo{pages}{014105}
  (\bibinfo{year}{2007}).

\bibitem[{\citenamefont{Beaudet et~al.}(2008)\citenamefont{Beaudet, Casula,
  Kim, Sorella, and Martin}}]{beaudet_molecular_2008}
\bibinfo{author}{\bibfnamefont{T.~D.} \bibnamefont{Beaudet}},
  \bibinfo{author}{\bibfnamefont{M.}~\bibnamefont{Casula}},
  \bibinfo{author}{\bibfnamefont{J.}~\bibnamefont{Kim}},
  \bibinfo{author}{\bibfnamefont{S.}~\bibnamefont{Sorella}}, \bibnamefont{and}
  \bibinfo{author}{\bibfnamefont{R.~M.} \bibnamefont{Martin}},
  \bibinfo{journal}{Journal of Chemical Physics}
  \textbf{\bibinfo{volume}{129}}, \bibinfo{pages}{164711}
  (\bibinfo{year}{2008}).

\bibitem[{\citenamefont{Santra et~al.}(2008)\citenamefont{Santra, Michaelides,
  Fuchs, Tkatchenko, Filippi, and Scheffler}}]{santra_accuracy_2008}
\bibinfo{author}{\bibfnamefont{B.}~\bibnamefont{Santra}},
  \bibinfo{author}{\bibfnamefont{A.}~\bibnamefont{Michaelides}},
  \bibinfo{author}{\bibfnamefont{M.}~\bibnamefont{Fuchs}},
  \bibinfo{author}{\bibfnamefont{A.}~\bibnamefont{Tkatchenko}},
  \bibinfo{author}{\bibfnamefont{C.}~\bibnamefont{Filippi}}, \bibnamefont{and}
  \bibinfo{author}{\bibfnamefont{M.}~\bibnamefont{Scheffler}},
  \bibinfo{journal}{Journal of Chemical Physics}
  \textbf{\bibinfo{volume}{129}}, \bibinfo{pages}{194111}
  (\bibinfo{year}{2008}).

\bibitem[{\citenamefont{Ma et~al.}(2009)\citenamefont{Ma, Alfe, Michaelides,
  and Wang}}]{ma_water-benzene_2009}
\bibinfo{author}{\bibfnamefont{J.}~\bibnamefont{Ma}},
  \bibinfo{author}{\bibfnamefont{D.}~\bibnamefont{Alfe}},
  \bibinfo{author}{\bibfnamefont{A.}~\bibnamefont{Michaelides}},
  \bibnamefont{and} \bibinfo{author}{\bibfnamefont{E.}~\bibnamefont{Wang}},
  \bibinfo{journal}{Journal of Chemical Physics}
  \textbf{\bibinfo{volume}{130}}, \bibinfo{pages}{154303}
  (\bibinfo{year}{2009}).

\bibitem[{\citenamefont{Umrigar et~al.}(2007)\citenamefont{Umrigar, Toulouse,
  Filippi, Sorella, and Hennig}}]{umrigar_alleviation_2007}
\bibinfo{author}{\bibfnamefont{C.~J.} \bibnamefont{Umrigar}},
  \bibinfo{author}{\bibfnamefont{J.}~\bibnamefont{Toulouse}},
  \bibinfo{author}{\bibfnamefont{C.}~\bibnamefont{Filippi}},
  \bibinfo{author}{\bibfnamefont{S.}~\bibnamefont{Sorella}}, \bibnamefont{and}
  \bibinfo{author}{\bibfnamefont{R.~G.} \bibnamefont{Hennig}},
  \bibinfo{journal}{Physical Review Letters} \textbf{\bibinfo{volume}{98}},
  \bibinfo{pages}{110201} (\bibinfo{year}{2007}).

\bibitem[{\citenamefont{Bajdich et~al.}(2010)\citenamefont{Bajdich, Tiago,
  Hood, Kent, and Reboredo}}]{bajdich_systematic_2010}
\bibinfo{author}{\bibfnamefont{M.}~\bibnamefont{Bajdich}},
  \bibinfo{author}{\bibfnamefont{M.~L.} \bibnamefont{Tiago}},
  \bibinfo{author}{\bibfnamefont{R.~Q.} \bibnamefont{Hood}},
  \bibinfo{author}{\bibfnamefont{P.~R.~C.} \bibnamefont{Kent}},
  \bibnamefont{and} \bibinfo{author}{\bibfnamefont{F.~A.}
  \bibnamefont{Reboredo}}, \bibinfo{journal}{Physical Review Letters} \textbf{\bibinfo{volume}{104}},
  \bibinfo{pages}{193001} 
   (\bibinfo{year}{2010}).

\bibitem[{\citenamefont{Grossman}(2002)}]{grossman_benchmark_2002}
\bibinfo{author}{\bibfnamefont{J.~C.} \bibnamefont{Grossman}},
  \bibinfo{journal}{Journal of Chemical Physics}
  \textbf{\bibinfo{volume}{117}}, \bibinfo{pages}{1434} (\bibinfo{year}{2002}).

\bibitem[{\citenamefont{Nemec et~al.}(2010)\citenamefont{Nemec, Towler, and
  Needs}}]{nemec_benchmark_2010}
\bibinfo{author}{\bibfnamefont{N.}~\bibnamefont{Nemec}},
  \bibinfo{author}{\bibfnamefont{M.~D.} \bibnamefont{Towler}},
  \bibnamefont{and} \bibinfo{author}{\bibfnamefont{R.~J.} \bibnamefont{Needs}},
  \bibinfo{journal}{Journal of Chemical Physics}
  \textbf{\bibinfo{volume}{132}}, \bibinfo{pages}{034111}
  (\bibinfo{year}{2010}).

\bibitem[{\citenamefont{Mitáš et~al.}(1991)\citenamefont{Mitáš, Shirley,
  and Ceperley}}]{mitas_nonlocal_1991}
\bibinfo{author}{\bibfnamefont{L.}~\bibnamefont{Mitáš}},
  \bibinfo{author}{\bibfnamefont{E.~L.} \bibnamefont{Shirley}},
  \bibnamefont{and} \bibinfo{author}{\bibfnamefont{D.~M.}
  \bibnamefont{Ceperley}}, \bibinfo{journal}{Journal of Chemical Physics}
  \textbf{\bibinfo{volume}{95}}, \bibinfo{pages}{3467} (\bibinfo{year}{1991}).

\bibitem[{\citenamefont{Burkatzki et~al.}(2007)\citenamefont{Burkatzki,
  Filippi, and Dolg}}]{burkatzki_energy-consistent_2007}
\bibinfo{author}{\bibfnamefont{M.}~\bibnamefont{Burkatzki}},
  \bibinfo{author}{\bibfnamefont{C.}~\bibnamefont{Filippi}}, \bibnamefont{and}
  \bibinfo{author}{\bibfnamefont{M.}~\bibnamefont{Dolg}},
  \bibinfo{journal}{Journal of Chemical Physics}
  \textbf{\bibinfo{volume}{126}}, \bibinfo{pages}{234105}
  (\bibinfo{year}{2007}).

\bibitem[{gam()}]{gamess09}
\bibinfo{note}{GAMESS Version 12 JAN 2009 (R1) From Iowa State University. M.W.
  Schmidt, K.K. Baldridge, J.A. Boatz, S.T. Elbert, M.S. Gordon, J.H. Jensen,
  S. Koseki, N. Matsunaga, K.A. Nguyen, S.J. Su, T.L. Windus, M. Dupuis, J.A.
  Montgomery, J. Comput. Chem. {\bf 14}, 1347-1363 (1993).}

\bibitem[{\citenamefont{Wagner et~al.}(2009)\citenamefont{Wagner, Bajdich, and
  Mitas}}]{wagner_qwalk_2009}
\bibinfo{author}{\bibfnamefont{L.~K.} \bibnamefont{Wagner}},
  \bibinfo{author}{\bibfnamefont{M.}~\bibnamefont{Bajdich}}, \bibnamefont{and}
  \bibinfo{author}{\bibfnamefont{L.}~\bibnamefont{Mitas}},
  \bibinfo{journal}{Journal of Computational Physics}
  \textbf{\bibinfo{volume}{228}}, \bibinfo{pages}{3390} (\bibinfo{year}{2009}).

\bibitem[{\citenamefont{Becke}(1988)}]{becke_density-functional_1988}
\bibinfo{author}{\bibfnamefont{A.~D.} \bibnamefont{Becke}},
  \bibinfo{journal}{Physical Review A} \textbf{\bibinfo{volume}{38}},
  \bibinfo{pages}{3098} (\bibinfo{year}{1988}).

\bibitem[{\citenamefont{Perdew and
  Zunger}(1981)}]{perdew_self-interaction_1981}
\bibinfo{author}{\bibfnamefont{J.~P.} \bibnamefont{Perdew}} \bibnamefont{and}
  \bibinfo{author}{\bibfnamefont{A.}~\bibnamefont{Zunger}},
  \bibinfo{journal}{Physical Review B} \textbf{\bibinfo{volume}{23}},
  \bibinfo{pages}{5048} (\bibinfo{year}{1981}).

\bibitem[{\citenamefont{Perdew et~al.}(1996)\citenamefont{Perdew, Burke, and
  Ernzerhof}}]{perdew_generalized_1996}
\bibinfo{author}{\bibfnamefont{J.~P.} \bibnamefont{Perdew}},
  \bibinfo{author}{\bibfnamefont{K.}~\bibnamefont{Burke}}, \bibnamefont{and}
  \bibinfo{author}{\bibfnamefont{M.}~\bibnamefont{Ernzerhof}},
  \bibinfo{journal}{Physical Review Letters} \textbf{\bibinfo{volume}{77}},
  \bibinfo{pages}{3865} (\bibinfo{year}{1996}).

\bibitem[{\citenamefont{Ernzerhof and
  Scuseria}(1999)}]{ernzerhof_assessment_1999}
\bibinfo{author}{\bibfnamefont{M.}~\bibnamefont{Ernzerhof}} \bibnamefont{and}
  \bibinfo{author}{\bibfnamefont{G.~E.} \bibnamefont{Scuseria}},
  \bibinfo{journal}{Journal of Chemical Physics}
  \textbf{\bibinfo{volume}{110}}, \bibinfo{pages}{5029} (\bibinfo{year}{1999}).

\bibitem[{\citenamefont{Adamo and Barone}(1999)}]{adamo_toward_1999}
\bibinfo{author}{\bibfnamefont{C.}~\bibnamefont{Adamo}} \bibnamefont{and}
  \bibinfo{author}{\bibfnamefont{V.}~\bibnamefont{Barone}},
  \bibinfo{journal}{Journal of Chemical Physics}
  \textbf{\bibinfo{volume}{110}}, \bibinfo{pages}{6158} (\bibinfo{year}{1999}).

\bibitem[{\citenamefont{Kummel and
  Kronik}(2008)}]{kummel_orbital-dependent_2008}
\bibinfo{author}{\bibfnamefont{S.}~\bibnamefont{Kummel}} \bibnamefont{and}
  \bibinfo{author}{\bibfnamefont{L.}~\bibnamefont{Kronik}},
  \bibinfo{journal}{Reviews of Modern Physics} \textbf{\bibinfo{volume}{80}},
  \bibinfo{pages}{3} (\bibinfo{year}{2008}).

\end{thebibliography}

\end{document}